\documentclass[aps,twocolumn,pre,floatfix]{revtex4-2}

\usepackage{xcolor,hyperref}
\hypersetup{
   colorlinks,
   linkcolor={blue!50!black},
   citecolor={blue!50!black},
   urlcolor={blue!80!black}
}

\usepackage{epsfig, amsmath}
\usepackage{bm}
\usepackage{soul,ulem}
\usepackage{hyperref}
\usepackage{footmisc}
\usepackage{wrapfig}
\usepackage{amssymb}
\DeclareMathAlphabet{\mathitbf}{OML}{cmm}{b}{it}

\renewcommand{\=}{\!=\!}

\newcommand{\uv}{\mathitbf u}

\newcommand{\kv}{\mathitbf k}

\newcommand{\av}{\mathitbf a}
\newcommand{\rv}{\mathitbf r}
\newcommand{\nv}{\mathitbf n}

\newcommand{\Psiv}{\bm{\Psi}}

\newcommand{\sFrac}[2]{{\textstyle\frac{#1}{#2}}}
\newcommand{\dbar}{{\,\mathchar'26\mkern-12mu d}}

\renewcommand{\=}{\!=\!}

\definecolor{darkGreen}{RGB}{4,161,85}

\begin{document}

\title{Enumerating low-frequency nonphononic vibrations in computer glasses}
\author{Edan Lerner$^{1}$}
\email{Corresponding author: e.lerner@uva.nl}
\author{Avraham Moriel$^{2}$}
\author{Eran Bouchbinder$^{3}$}
\email{eran.bouchbinder@weizmann.ac.il}
\affiliation{$^{1}$Institute of Theoretical Physics, University of Amsterdam, Science Park 904, 1098 XH Amsterdam, the Netherlands\\
$^{2}$Dept.~of Mechanical \& Aerospace Engineering, Princeton University, Princeton, New Jersey 08544, USA\\
$^{3}$Chemical and Biological Physics Department, Weizmann Institute of Science, Rehovot 7610001, Israel}

\begin{abstract}
In addition to Goldstone phonons that generically emerge in the low-frequency vibrational spectrum of any solid, crystalline or glassy, structural glasses also feature other low-frequency vibrational modes. The nature and statistical properties of these modes --- often termed `excess modes' --- have been the subject of decades-long investigation. Studying them, even using well-controlled computer glasses, has proven challenging due to strong spatial hybridization effects between phononic and nonphononic excitations, which hinder quantitative analyses of the nonphononic contribution ${\cal D}_{\rm G}(\omega)$ to the total spectrum ${\cal D}(\omega)$, per frequency $\omega$. Here, using recent advances indicating that ${\cal D}_{\rm G}(\omega)\!=\!{\cal D}(\omega)-{\cal D}_{\rm D}(\omega)$, where ${\cal D}_{\rm D}(\omega)$ is Debye's spectrum of phonons, we present a simple and straightforward scheme to enumerate nonphononic modes in computer glasses. Our analysis establishes that nonphononic modes in computer glasses indeed make an additive contribution to the total spectrum, including in the presence of strong hybridizations. Moreover, it cleanly reveals the universal ${\cal D}_{\rm G}(\omega)\!\sim\!\omega^4$ tail of the nonphononic spectrum, and opens the way for related analyses of experimental spectra of glasses.
\end{abstract}

\maketitle

\section{Introduction}

The Hamiltonian of solids --- glassy, crystalline or otherwise --- generically features translational invariance, and therefore low-frequency waves (long wavelength phonons) populate the vibrational spectra of solids, independently of their microscopic structure~\cite{kittel2005introduction}. In structural glasses formed by cooling a melt, additional, nonphononic low-energy excitations emerge due to the existence of structural frustration and disorder~\cite{JCP_Perspective}. These excitations are quasilocalized in space, i.e., they feature a disordered core of linear size of 5-10 interparticle distances, decorated by far fields that decay at distance $r$ away from the core as $r^{-(\dbar-1)}$, in $\dbar$ spatial dimensions~\cite{JCP_Perspective}.

Quasilocalized excitations (QLEs) play important roles in controlling the mechanical~\cite{micromechanics2016,david_fracture_mrs_2021} and thermodynamic~\cite{soft_potential_model_1991} properties of glassy solids. They have been shown to follow a universal $\sim\!\omega^4$ nonphononic spectrum~\cite{modes_prl_2016,ikeda_pnas} in the low-frequencies $\omega\!\to\!0$ limit~\cite{footnote1} in a broad class of computer models of disordered solids~\cite{modes_prl_2020,disordered_crystals_prl_2022,tommaso_arXiv_2024}, independently of microscopic details~\cite{universal_VDoS_ip}, spatial dimension~\cite{modes_prl_2018,2d_spectra_jcp_2022,Atsushi_high_d_pre_2020,atsushi_2d_pinning_jcp_2023}, and the nonequilibrium formation history/protocol~\cite{LB_modes_2019,pinching_pnas}.
\begin{figure}[ht!]
 \includegraphics[width = 0.5\textwidth]{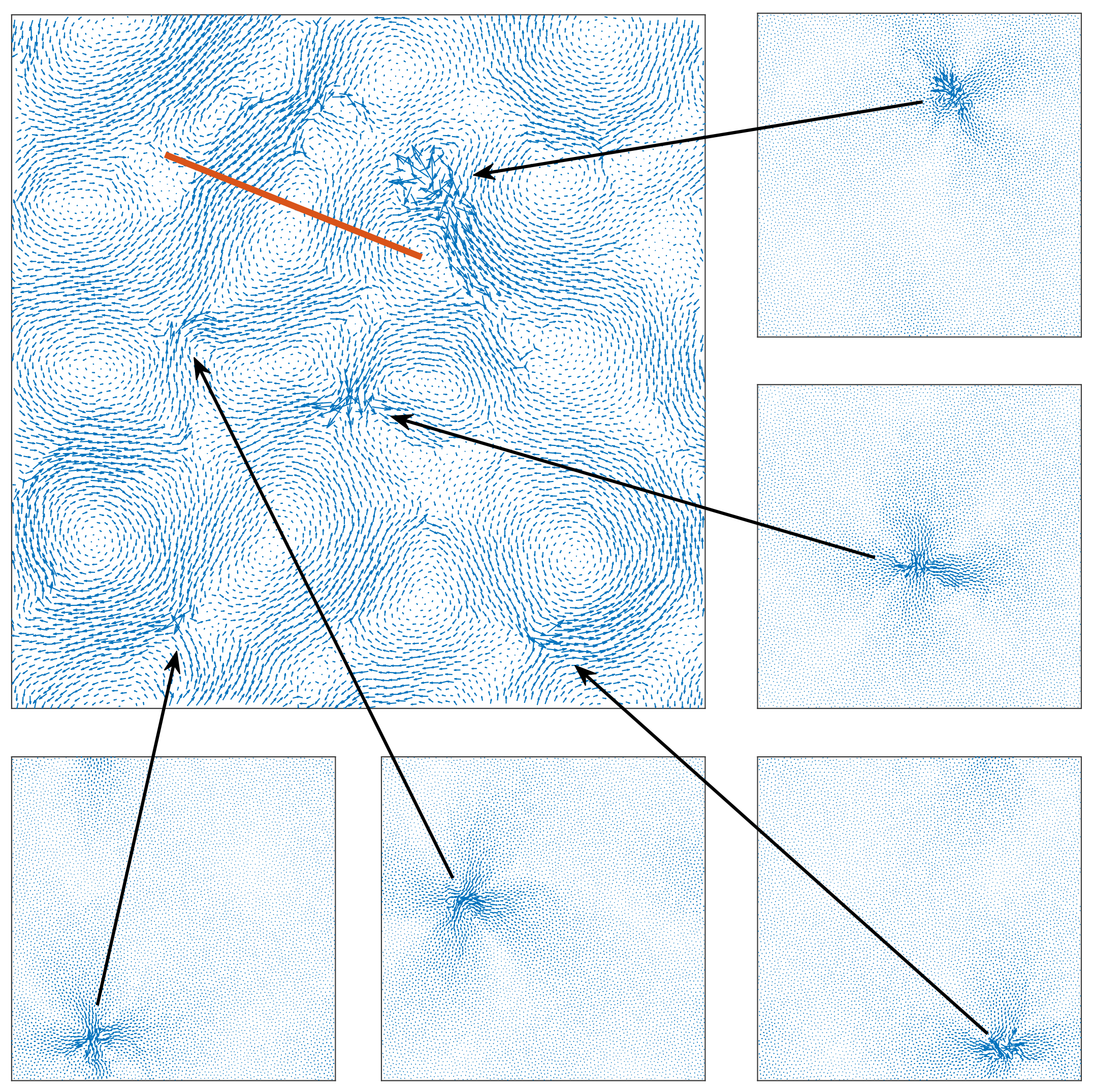}
 \vspace{-0.4cm}
  \caption{\footnotesize (large panel) A vibrational mode in a 2D computer glass of $N\!=\!8100$ particles at frequency $\omega\!=\!0.037\,\omega_{\rm D}$ ($\omega_{\rm D}$ is Debye's frequency~\cite{footnote3}, and see~\cite{boson_peak_2d_jcp_2023} for model details), which is comprised of several hybridized QLEs with phonons. The red bar indicates the wavelength of phonons having the same frequency $\omega$ as that of the vibrational mode. (smaller panels) 5 QLEs that were isolated and dehybridized as explained in~\cite{boson_peak_2d_jcp_2023}, see text for additional discussion (each panel corresponds to the entire system shown in the large panel).}
\label{fig:introduction_fig}
\end{figure}

A major obstacle hindering a direct and clean investigation of the statistical-mechanical properties of QLEs in computer glasses, including their nonphononic vibrational density of states (VDoS) ${\cal D}_{\rm G}(\omega)$, is that they only assume the form of harmonic vibrational modes (normal modes) when their vibrational frequencies $\omega$ fall below or in between quantized phonon bands~\cite{phonon_widths} in finite-size glass samples. The range of frequencies $\omega$ for which these conditions are met shrinks with the size of the glass (quantified by the number of particles $N$ or the linear size $L\!\sim\!N^{1/3}$ in three dimensions)~\cite{footnote2}, such that in large systems low-frequency vibrations consist of spatially hybridized (mixed) QLEs and phonons, up to and in the vicinity of the boson peak frequency, as demonstrated in Fig.~\ref{fig:introduction_fig} and in~\cite{boson_peak_2d_jcp_2023}. At the same time, there is a need to study large computer glasses since the nonphononic VDoS may feature strong finite-size effects in smaller samples~\cite{lerner2019finite,2d_spectra_jcp_2022,Xu_modes_nat_comm_2024}. These challenges led to the development of various nonlinear methods to dehybridize QLEs and phonons~\cite{SciPost2016,manning_defects,episode_1_2020,pseudo_harmonic_prl,david_detecting_qles_pre_2023}, having measurable success in shedding light on the statistical properties of QLEs~\cite{cge_paper,JCP_Perspective,boson_peak_2d_jcp_2023}. However, these methods are not yet fully exhaustive, and their numerical implementation might be cumbersome and computationally expensive.

Recently, it has been suggested~\cite{boson_peak_2d_jcp_2023,moriel2024boson} that while nonphononic excitations generically hybridize (mix) with phonons in space, their number per frequency $\omega$ contributes additively to the total ${\cal D}(\omega)$. That is, it was suggested that the latter follows an additive structure of the form
\begin{equation}
{\cal D}(\omega) = {\cal D}_{\rm G}(\omega)+{\cal D}_{\rm D}(\omega) \ ,
\label{eq:additive}
\end{equation}
where ${\cal D}_{\rm D}(\omega)$ is Debye's VDoS of phonons. Similar suggestions were made earlier in~\cite{YANNOPOULOS20064541,KALAMPOUNIAS20064619,Schirmacher_prl_2007,grzegorz_2d_modes_prl_2021} and some preliminary support to Eq.~\eqref{eq:additive} was provided in~\cite{boson_peak_2d_jcp_2023,moriel2024boson}. In this work, we directly validate Eq.~\eqref{eq:additive} and in so doing we develop a simple and straightforward approach to enumerate low-frequency nonphononic vibrational modes in computer glasses. The approach developed here, which is widely applicable to any computer glass, is useful in determining the scaling regime of the universal ${\cal D}_{\rm G}(\omega)\!\sim\!\omega^4$ tail of the nonphononic spectrum, allows to overcome the aforementioned finite-size effects in the nonphononic spectrum, and also opens the way for related analyses of experimental spectra of glasses based on Eq.~\eqref{eq:additive}.

The paper is organized as follows; in Sect.~\ref{sec:total}, we describe the model systems considered, the observables measured and the numerical methods employed. In Sect.~\ref{sec:debye}, we briefly recap Debye's theory of phonons in homogeneous, isotropic media, allowing to calculate ${\cal D}_{\rm D}(\omega)$. Section~\ref{sec:results} presents our enumeration analysis of excess vibrational modes in computer glasses based on Eq.~\eqref{eq:additive}. We summarize our work and highlight future research directions in Sect.~\ref{sec:summary}.

\section{Methodology}
\label{sec:methods}

Equation~\eqref{eq:additive} implies that the nonphononic VDoS is given by ${\cal D}_{\rm G}(\omega)\!=\!{\cal D}(\omega)-{\cal D}_{\rm D}(\omega)$. Consequently, one needs to generate computer glasses, measure their total VDoS ${\cal D}(\omega)$, compute Debye's VDoS of phonons ${\cal D}_{\rm D}(\omega)$ based on the measured elastic properties of the glass and finally evaluate thier difference to obtain ${\cal D}_{\rm G}(\omega)$. In this section, we explain how this is done.

\subsection{Computer glasses and the total VDoS ${\cal D}(\omega)$}
\label{sec:total}

We employ two simple models of structural glasses in three dimensions (3D) in which point-like particles interact via an inverse power-law $\sim\!r^{-10}$ pairwise potential, with $r$ denoting the pairwise distance. The first model --- referred to in what follows as BIPL --- is a 50:50 binary mixture of two types of particles with different effective sizes, whereas the second model is a polydisperse glass former referred to in what follows as PIPL. Details about the potential and parameters employed can be found in~\cite{cge_paper} for the BIPL model, and in~\cite{boring_paper} for the PIPL model. Ensembles of computer glasses were created by a continuous quench from high-temperature liquid states into the glass phase at a rate $\dot{T}\!=\!10^{-3}$ for the BIPL system (in simulational units, see~\cite{cge_paper}), whereas in the PIPL we instantaneously quenched glasses from equilibrium liquid states at parent temperature $T_{\rm p}\!=\!0.32$ (in simulational units, see~\cite{boring_paper}). Instantaneous quenches were performed using a conventional conjugate gradient method. The total VDoS ${\cal D}(\omega)$ was calculated using ARPACK~\cite{arpack}. Elastic moduli, which are needed to obtain Debye's phononic VDoS ${\cal D}_{\rm D}(\omega)$ next, were calculated using the microscopic expressions as provided, e.g.,~in~\cite{boring_paper}.

\subsection{Debye's phononic VDoS ${\cal D}_{\rm D}(\omega)$}
\label{sec:debye}

We consider isotropic, homogeneous materials of linear size $L$ in $\dbar$ dimensions, characterized by shear modulus $G$, bulk modulus $K$ and mass density $\rho$. At the continuum level, under the assumption of small deformations, the displacement field $\uv(\rv,t)$ is governed by the Navier-Lam\'e equation~\cite{landau_lifshitz_elasticity}
\begin{equation}
\rho\,\ddot{\uv} = (K\!+\!G/\dbar)\,\nabla(\nabla\cdot\uv) + G\,\nabla^2\uv\ .
\end{equation}
Plane-waves (long wavelength phonons) of the form
\begin{equation}
    \Psiv_{{\rm s},\ell}(\rv) \propto \hat{\av}_{{\rm s},\ell}\,e^{i\kv\cdot\rv}
\end{equation}
correspond to oscillating solutions, $\ddot{\Psiv}\!\propto\!-\Psiv$. Here, $\kv$ in the wavevector and $\hat{\av}_{{\rm s},\ell}$ are polarization vectors that satisfy  $\hat{\av}_{\ell}\cdot\kv/|\kv|\!=\!1$ for sound (longitudinal) waves and $\hat{\av}_{\rm s}\cdot\kv/|\kv|\!=\!0$ for $\dbar\!-\!1$ orthogonal shear waves.

Under periodic boundary conditions, the wavevector $\kv$ assumes the form
\begin{equation}\label{eq:wavevector}
    \kv = \frac{2\pi}{L}\nv\,,
\end{equation}
where $\nv$ is a $\dbar$-dimensional vector of integers. The oscillation frequency of these long wavelength phonons is given by $\omega\!=\!c_{\rm s}|\kv|$ for shear waves, with $c_{\rm s}\!\equiv\!\sqrt{G/\rho}$, or $\omega\!=\!c_{\ell}|\kv|$ for sound waves, with $c_{\ell}\!\equiv\!\sqrt{(K\!+\!(2\!-\!2/\dbar)G)/\rho}$.

\begin{figure*}[ht!]
 \includegraphics[width = 1.0\textwidth]{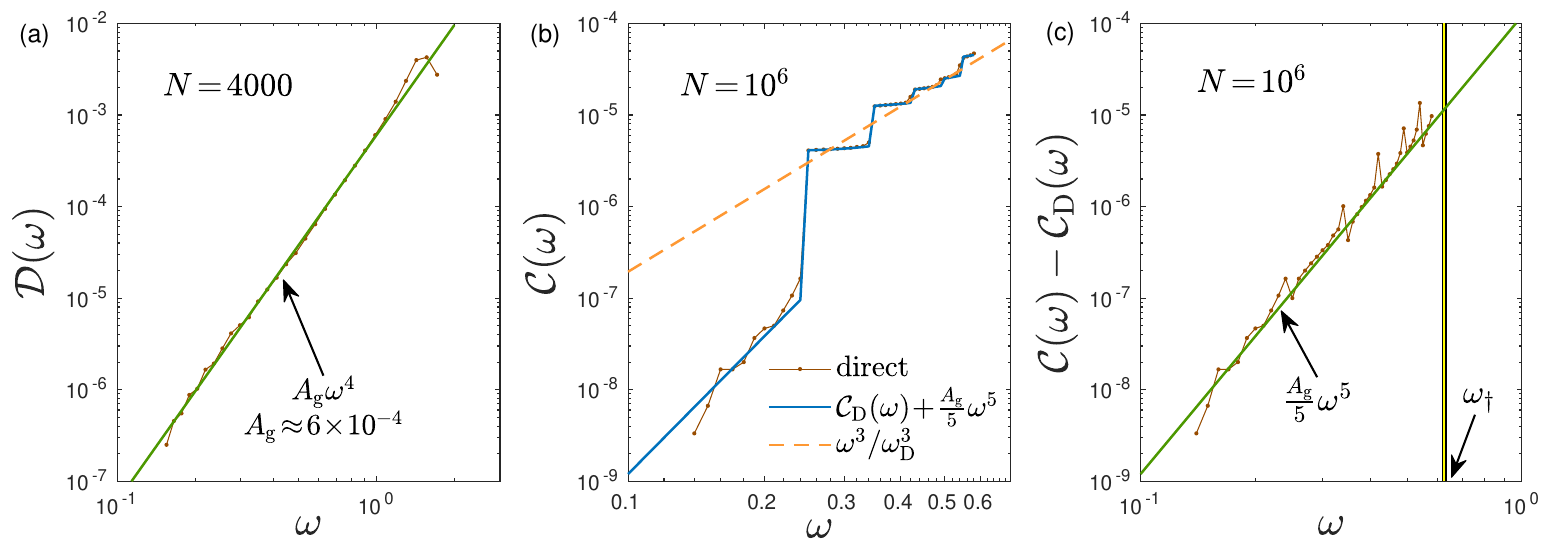}
 \vspace{-0.4cm}
  \caption{\footnotesize (a) The low-frequency tail of the VDoS ${\cal D}(\omega)$ averaged over $10^5$ independently formed BIPL computer glasses of $N\!=\!4000$ particles. It is fitted to the quartic law $A_{\rm g}\,\omega^4$ to extract $A_{\rm g}$ (as indicated). (b) The low-frequency cumulative VDoS ${\cal C}(\omega)$ for the same computer glasses, this time with 100 realizations of $N\!=\!10^6$ particles. Using $A_{\rm g}$ measured in panel (a), we superimpose the sum ${\cal C}_{\rm D}(\omega)\!+\!A_{\rm g}\omega^5/5$ to find very good agreement, see text for additional discussion. The dashed line represents ${\cal C}_{\rm D}^{(L\to\infty)}(\omega)\!=\!\omega^3/\omega_{\rm D}^3$. (c) ${\cal C}(\omega)-{\cal C}_{\rm D}(\omega)$ vs.~$\omega$ on a double-logarithmic scale (dotted line), corresponding to the same data shown in panel (b). The superimposed solid line corresponds to ${\cal C}_{\rm G}(\omega)\!=\!A_{\rm g}\omega^5/5$, with $A_{\rm g}$ extracted in panel (a). The vertical line corresponds to the crossover frequency $\omega_\dagger(L)$. See text for the discussion of the results.}
\label{fig:extract_Ag}
\end{figure*}

The VDoS ${\cal D}_{\rm D}(\omega)$ of these long wavelength phonons corresponds to the number of orthogonal solutions $\Psiv$ in a narrow frequency range around $\omega$. For macroscopically large systems, i.e., in the thermodynamic $L\!\to\!\infty$ limit, the spectrum is continuous and follows Debye's theory
\begin{equation}
{\cal D}_{\rm D}(\omega)= A_{\rm D}\,\omega^{\dbar-1} = \frac{\dbar}{\omega_{\rm D}^\dbar}\,\omega^{\dbar-1} \ ,
\label{eq:Debye_VDOS}
\end{equation}
where $A_{\rm D}\=\dbar/\omega_{\rm D}^\dbar$ is Debye's prefactor and $\omega_{\rm D}$ is Debye's frequency in $\dbar$ dimensions. In 3D ($\dbar\=3$), which we focus on hereafter, one has
\begin{equation}
    \label{eq:debye_freq}
    \omega_{\rm D} \equiv \left(
    \frac{18\pi^2(N/L^3)}{2c_{\rm s}^{-3} + c_\ell^{-3}}
    \right)^{1/3} \ ,
\end{equation}
where the corresponding expression in 2D ($\dbar\=2$) is given in~\cite{footnote3} (see also the caption of Fig.~\ref{fig:introduction_fig}). Finally, we note that Debye's cumulative VDoS (the number of long wavelength phonons having frequencies $\le\!\omega$) takes the form ${\cal C}_{\rm D}^{(L\to\infty)}(\omega)\=\omega^3/\omega_{\rm D}^3$ in 3D. The superscript `$(L\!\to\!\infty)$' serves to distinguish the above expression from its finite $L$ counterpart (recall that $L\!\sim\!N^{1/3}$), to be considered below.

A preliminary application of Eq.~\eqref{eq:additive} to experimental data of boron oxide glasses~\cite{moriel2024boson} invoked Eq.~\eqref{eq:Debye_VDOS} with an experimentally determined Debye's prefactor $A_{\rm D}$, as the laboratory glasses were macroscopically large, $L\!\to\!\infty$. When applying Eq.~\eqref{eq:Debye_VDOS} to computer glasses of finite linear size $L$, the low frequency spectrum becomes quantized, i.e., composed of discrete phonon bands. Consequently, Debye's continuous VDoS of Eq.~\eqref{eq:Debye_VDOS} is no longer valid at the lowest frequencies $\omega$ in finite computer glasses, and one has to account for the emerging quantization.

In order to determine the finite $L$ counterpart of ${\cal C}_{\rm D}^{(L\to\infty)}(\omega)$, one needs to enumerate the phonons in each phonon band, appearing at discrete frequencies below $\omega$. In the absence of glassy (structural) disorder, the phonon bands are degenerated, i.e., multiple wavevectors feature the very same frequency. In glasses, the disorder lifts the degeneracy, resulting in a finite spectral width $\Delta\omega$ per band, which is well-understood theoretically~\cite{phonon_widths}. However, as we are only interested in the number of phonons per band, $\Delta\omega$ is of no importance here (see some additional discussion of this point below), and we focus on the degeneracy of each phonon band.

The degeneracy of phonon bands is obtained from the $\dbar$-dimensional integer sum-of-squares problem~\cite{phonon_widths}
\begin{equation}
\label{eq:integer_sum_of_squares}
\nv\cdot\nv = \sum_{i=1}^\dbar n_i^2 = q\,,
\end{equation}
where $q\!>\!0$ is an integer that labels phonon bands. The number ${\cal N}(q)$ of different solutions to Eq.~(\ref{eq:integer_sum_of_squares}) --- including permutations --- is defined as the degeneracy of the $q^{\mbox{\scriptsize th}}$ phonon band. The finite $L$, Debye's cumulative VDoS ${\cal C}_{\rm D}(\omega)$ is then given by
\begin{equation}
    {\cal C}_{\rm D}(\omega) = \frac{\displaystyle(\dbar-1)
    \sum_{q=1}^{q^{\mbox{\tiny max}}_{\rm s}(\omega)}{\cal N}(q) + \sum_{q=1}^{q^{\mbox{\tiny max}}_{\rm \ell}(\omega)}{\cal N}(q)
    }{N\dbar-\dbar}\,,
    \label{eq:cumulative_discrete}
\end{equation}
where $q_{{\rm s},\ell}^{\mbox{\tiny max}}(\omega)$ are the \emph{largest integers} that satisfy
\begin{equation}
    \sqrt{q_{{\rm s},\ell}^{\mbox{\tiny max}}(\omega)} \le \frac{\omega L}{2\pi c_{{\rm s},\ell}} \,.
\end{equation}

The resulting ${\cal C}_{\rm D}(\omega)$ is a \emph{step-wise} function, corresponding to the underlying discrete phonon bands, which may approach ${\cal C}_{\rm D}^{(L\to\infty)}(\omega)$ in a certain frequency range for a given $L$ (as demonstrated explicitly in Fig.~\ref{fig:extract_Ag}b). In what follows, we will employ the $L$-dependent, step-wise ${\cal C}_{\rm D}(\omega)$ of Eq.~\eqref{eq:cumulative_discrete} in applying Eq.~\eqref{eq:additive} to finite-size computer glasses. That is, we will use the simple and
straightforward scheme described above for enumerating the excess, nonphononic modes in finite-size computer glasses based on the cumulative (integrated) version of Eq.~\eqref{eq:additive}, ${\cal C}_{\rm G}(\omega)\!=\!{\cal C}(\omega)\!-\!{\cal C}_{\rm D}(\omega)$.

\section{results}
\label{sec:results}

Once the total and Debye's VDoSs are at hand, the nonphononic VDoS ${\cal D}_{\rm G}(\omega)$ (or its cumulative counterpart ${\cal C}_{\rm G}(\omega)$) is immediately obtained as ${\cal D}_{\rm G}(\omega)\!=\!{\cal D}(\omega)-{\cal D}_{\rm D}(\omega)$, {\bf if} Eq.~\eqref{eq:additive} is valid, i.e., if indeed nonphononic modes make an additive contribution to the total VDoS. To assess the validity of the latter, one needs an independent procedure to obtain ${\cal D}_{\rm G}(\omega)$. As mentioned above, it has been extensively demonstrated that ${\cal D}_{\rm G}(\omega)\!=\!A_{\rm g}\omega^4$ at low frequencies~\cite{JCP_Perspective}, where the prefactor of the universal quartic law, $A_{\rm g}$, can be extracted by thoughtfully selecting small computer glass samples~\cite{modes_prl_2016,phonon_widths}. Consequently, our strategy would be to determine $A_{\rm g}$ using small computer glass samples and then test the validity of Eq.~\eqref{eq:additive} in significantly larger samples.

We first apply this procedure to the BIPL model. In Fig.~\ref{fig:extract_Ag}a, we show the low-frequency tail of the VDoS ${\cal D}(\omega)$, measured in glassy samples of $N\!=\!4000$ particles. In this case, a substantial number of QLEs are observed below the lowest-frequency phonon band, allowing to reveal the ${\cal D}_{\rm G}(\omega)\!\sim\!\omega^4$ scaling. The lowest-frequency phonon band in this system resides at a frequency $2\pi c_{\rm s}/L\!\approx\!1.54$ (in simulational units), corresponding to the small bump seen at the high end of the $\sim\!\omega^4$ scaling regime in Fig.~\ref{fig:extract_Ag}a. This dataset allows us to cleanly extract the prefactor of the nonphononic VDoS ${\cal D}_{\rm G}(\omega)\!=\!A_{\rm g}\,\omega^4$, giving rise to $A_{\rm g}\!\approx\!6\!\times\!10^{-4}$ (in simulational units).
\begin{figure}[ht!]
 \includegraphics[width = 0.5\textwidth]{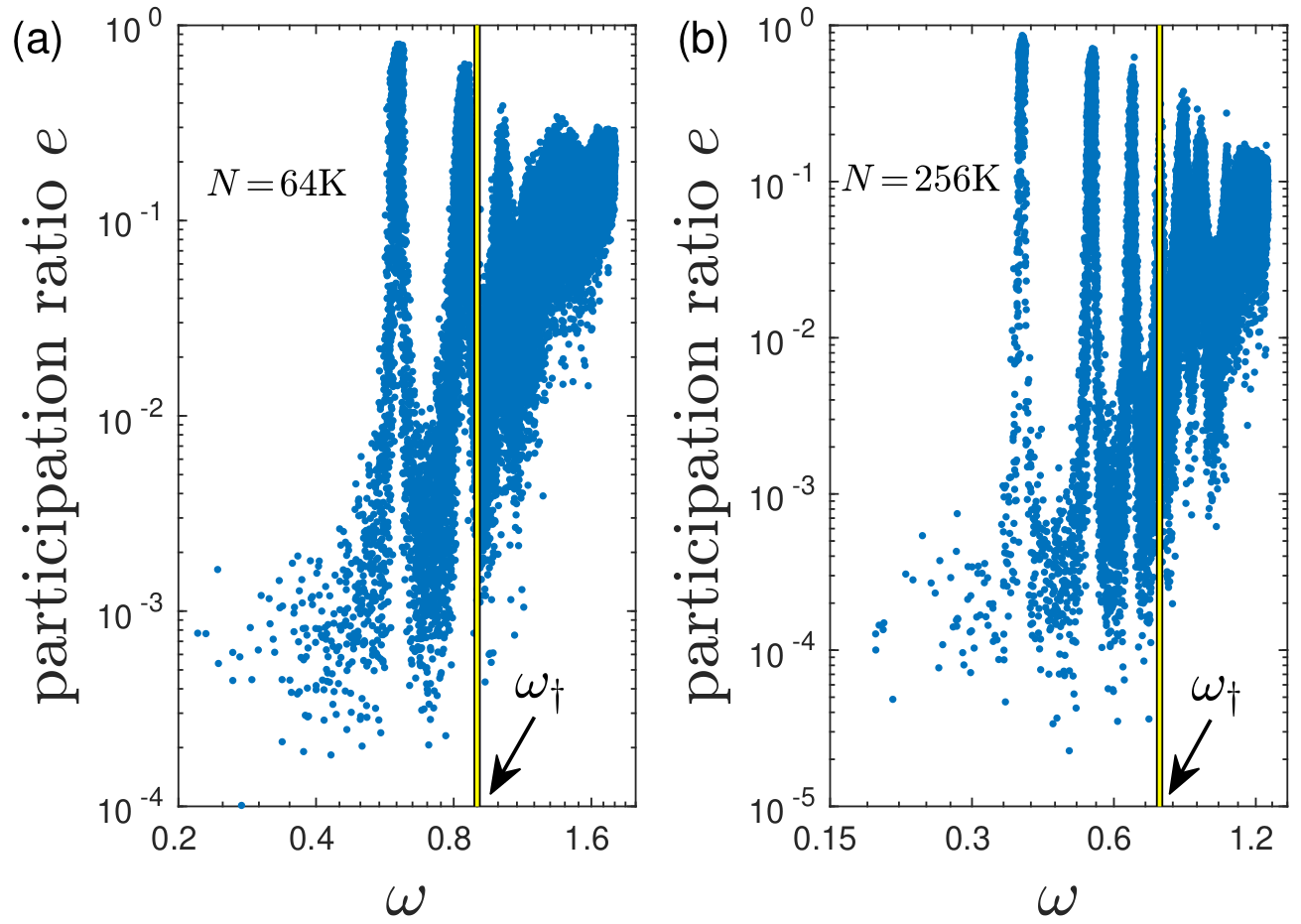}
 \vspace{-0.4cm}
  \caption{\footnotesize The participation ratio $e$ (see text for definition) calculated for individual vibrational modes is scatter-plotted against the mode's frequency $\omega$ for 3D BIPL glasses of (a) $N\!=\!64000$ and (b) $N\!=\!256000$ particles. The vertical lines mark the crossover frequency $\omega_\dagger(L)$ above which QLEs are typically hybridized in space, see text for discussion.}
\label{fig:omega_dagger_fig}
\end{figure}

\begin{figure*}[ht!]
 \includegraphics[width = 0.85\textwidth]{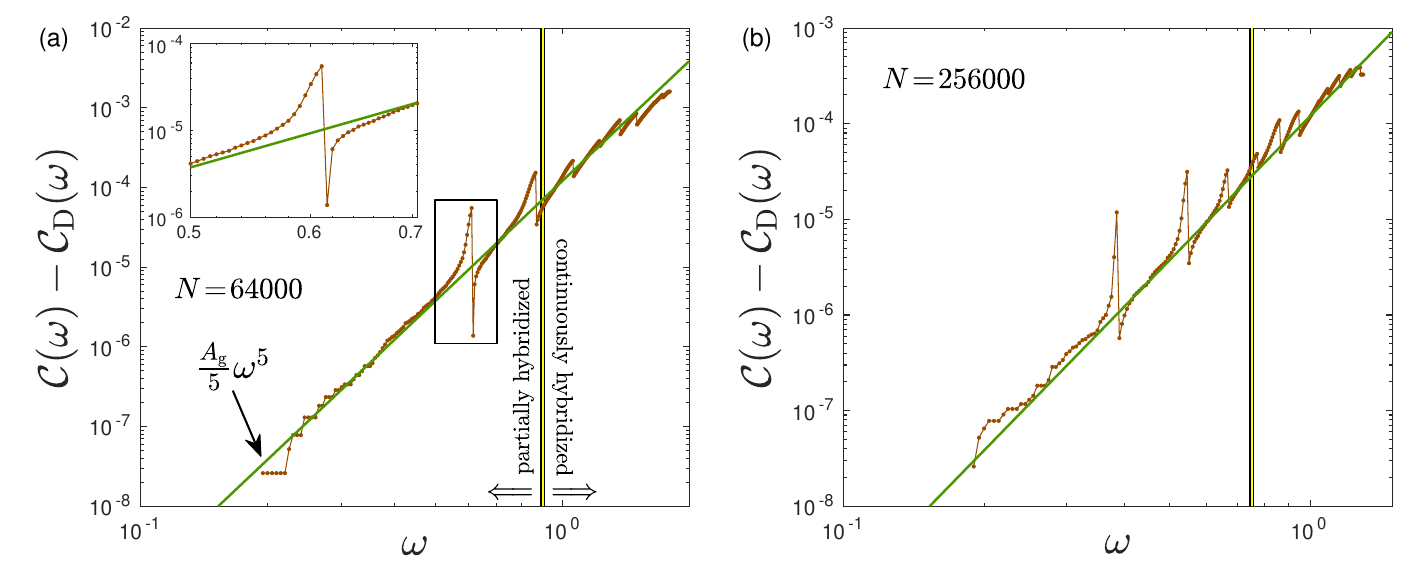}
 \vspace{-0.4cm}
  \caption{\footnotesize Enumerating nonphononic modes in computer glasses of different sizes (indicated in the legends) using the developed scheme. We follow the format of Fig.~\ref{fig:extract_Ag}c and plot ${\cal C}(\omega)\!-\!{\cal C}_{\rm D}(\omega)$ in dotted lines in panels (a) and (b). The vertical lines correspond to the crossover frequency $\omega_\dagger(L)$ (see Fig.~\ref{fig:omega_dagger_fig} and text for discussion). The continuous green lines correspond to the cumulative nonphononic VDoS ${\cal C}_{\rm G}(\omega)\!=\!A_{\rm g}\omega^5/5$, with $A_{\rm g}$ independently extracted in Fig.~\ref{fig:extract_Ag}a. The inset in panel (a) zooms in on the localized up and down deviations from the continuous line, which are marked by the rectangle in the main panel. See text for discussion.}
\label{fig:ipl_vdos}
\end{figure*}

With $A_{\rm g}$ at hand, we can now perform the first validity test of Eq.~\eqref{eq:additive}, which in turn allows us to accurately enumerate excess modes in larger systems. We present Eq.~\eqref{eq:additive} in its cumulative, low-frequency form as
\begin{equation}
    {\cal C}(\omega) = {\cal C}_{\rm D}(\omega) + \sFrac{1}{5}A_{\rm g}\omega^5\ ,
\label{eq:comulative_sum}
\end{equation}
where we used ${\cal C}_{\rm G}(\omega)\=A_{\rm g}\omega^5/5$ for small $\omega$ and computed the step-wise ${\cal C}_{\rm D}(\omega)$ as explained in Sect.~\ref{sec:debye}. In Fig.~\ref{fig:extract_Ag}b, we plot the directly measured cumulative VDoS ${\cal C}(\omega)$ for $N\!=\!10^6$ (i.e., nearly 3 order of magnitude larger than in panel (a)), and superimpose on it the sum ${\cal C}_{\rm D}(\omega)\!+\!A_{\rm g}\omega^5/5$, using the value of $A_{\rm g}$ extracted in panel (a). The step-wise contribution of ${\cal C}_{\rm D}(\omega)$ is apparent, and the agreement with the direct measurement is excellent. In addition, we superimpose ${\cal C}_{\rm D}^{(L\to\infty)}(\omega)\!=\!\omega^3/\omega_{\rm D}^3$ (see legend, following Eq.~\eqref{eq:Debye_VDOS}), demonstrating convergence to it above some frequency.

The results presented in Fig.~\ref{fig:extract_Ag}b do not yet support the validity of Eq.~\eqref{eq:comulative_sum}, and hence of Eq.~\eqref{eq:additive}, because they would have been the same for {\em every} nonphononic spectrum that satisfies ${\cal C}_{\rm G}(\omega)\!\ll\!{\cal C}_{\rm D}(\omega)$, not specifically for ${\cal C}_{\rm G}(\omega)\=A_{\rm g}\omega^5/5$. Consequently, we plot in Fig.~\ref{fig:extract_Ag}c the same data, but as ${\cal C}(\omega)-{\cal C}_{\rm D}(\omega)$ on a double-logarithmic scale and superimpose ${\cal C}_{\rm G}(\omega)\=A_{\rm g}\omega^5/5$, using $A_{\rm g}$ of Fig.~\ref{fig:extract_Ag}a. The observe agreement between ${\cal C}(\omega)-{\cal C}_{\rm D}(\omega)$ and ${\cal C}_{\rm G}(\omega)\=A_{\rm g}\omega^5/5$ is excellent, providing strong support to the additive structure of the VDoS of glasses in Eq.~\eqref{eq:additive}, and constituting one of our major results.

Figure~\ref{fig:extract_Ag} demonstrates that the nonphononic VDoS ${\cal D}_{\rm G}(\omega)\=A_{\rm g}\,\omega^4$ of QLEs can be simply and straightforwardly obtained in computer glasses using Eq.~\eqref{eq:additive}, despite the coexistence and spatial hybridization with phonons. The latter do have a signature in Fig.~\ref{fig:extract_Ag}c, in the form of localized deviations from the continuous ${\cal C}_{\rm G}(\omega)\=A_{\rm g}\omega^5/5$ line. These correspond to discrete phonon bands and manifest themselves because the bands are not strictly degenerate as in crystalline solids, but rather feature finite spectral widths $\Delta\omega$~\cite{phonon_widths}, as mentioned above. This issue is father discussed below, even though it has no implications for the extraction of ${\cal D}_{\rm G}(\omega)\=A_{\rm g}\,\omega^4$.

Clearly distinguished, discrete phonon bands exist up to an $L$-dependent crossover frequency scale $\omega_\dagger(L)\!\sim\!L^{-2/(\dbar+2)}$ (see ~\cite{phonon_widths,footnote2}), marked by the vertical line in Fig.~\ref{fig:extract_Ag}c. For $\omega\!>\!\omega_\dagger$, phonon bands are no longer distinguishable (they overlap one another, starting to merge and form a continuous spectrum) and QLEs are rather continuously hybridized with phonons (and with other QLEs, cf.~Fig.~\ref{fig:introduction_fig} above). The merging of phonon bands and the nearly continuous hybridization of QLEs with phonons are demonstrated in Fig.~\ref{fig:omega_dagger_fig}. In the figure, we scatter-plot the participation ratio $e\!\equiv\!\big[N\sum_i(\Psiv_i\cdot\Psiv_i)^2\big]^{-1}$ ($i$ denotes particle indices) of each individual vibrational mode $\Psiv$, calculated across 200 independently formed glass samples of $N\!=64000$ (panel (a)) and across 100 independently formed glass samples of $N\!=\!256000$ (panel (b)) particles. It is observed that for $\omega\!>\!\omega_\dagger$, discrete phonon bands are nearly indistinguishable and that the participation ratio $e$ attains values of ${\cal O}(1)$, indicative of spatial hybridization with extended phonons (compare to $e\!\sim\!{\cal O}(1/N)$ away from phonon bands for $\omega\!<\!\omega_\dagger$, and note the different $y$-axis range of the two panels).

Next, we ask whether Eq.~\eqref{eq:additive} remains valid also in the presence of stronger, more continuous hybridization between QLEs and phonons for $\omega\!>\!\omega_\dagger$. To address this question, we plot in Fig.~\ref{fig:ipl_vdos} ${\cal C}(\omega)\!-\!{\cal C}_{\rm D}(\omega)$ for the glass ensembles shown in Fig.~\ref{fig:omega_dagger_fig}, and superimpose ${\cal C}_{\rm G}(\omega)\=A_{\rm g}\omega^5/5$ in same format as in Fig.~\ref{fig:extract_Ag}c. The agreement between the two is excellent in both panels, both below and above the crossover frequency $\omega_\dagger(L)$, yet again supporting the additivity of the phononic and nonphononic VDoSs in Eq.~\eqref{eq:additive}. In the inset of Fig.~\ref{fig:ipl_vdos}a, we zoom in on the localized up and down deviations from the continuous ${\cal C}_{\rm G}(\omega)\=A_{\rm g}\omega^5/5$ line marked by the rectangle in the main panel. The inset clearly reveals that the up and down deviations, of spectral span determined by the corresponding width $\Delta\omega$ of the disorder-broadened phonon band~\cite{phonon_widths}, cancel out such that ${\cal C}(\omega)\!-\!{\cal C}_{\rm D}(\omega)$ returns to the ${\cal C}_{\rm G}(\omega)\=A_{\rm g}\omega^5/5$ line.

\begin{figure}[ht!]
 \includegraphics[width = 0.5\textwidth]{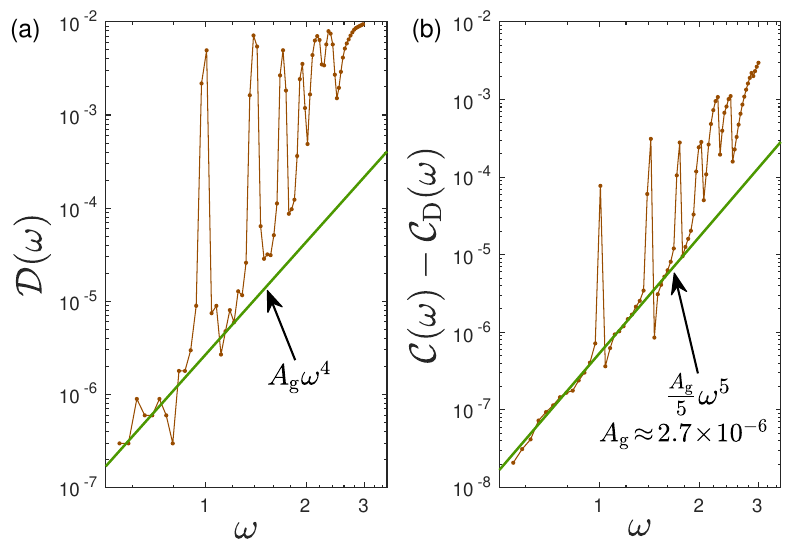}
 \vspace{-0.4cm}
  \caption{\footnotesize Enumerating nonphononic modes in ultrastable 3D PIPL computer glasses using the developed scheme (panel (b)), compared to the conventional approach employed in the literature (panel (a)). (a) The low-frequency VDoS ${\cal D}(\omega)$ (dotted line), revealing that QLEs are too scarce below the first phonon band in order to robustly extract the prefactor $A_{\rm g}$ of the tail of the nonphononic VDoS, ${\cal D}_{\rm G}(\omega)\!=\!A_{\rm g}\,\omega^4$. The continuous green line is obtained from  the analysis of panel (b), explained next. (b) ${\cal C}(\omega)\!-\!{\cal C}_{\rm D}(\omega)$ vs.~$\omega$ (dotted line), with a superimposed fit to ${\cal C}_{\rm G}(\omega)\!=\!A_{\rm g}\omega^5/5$ (continuous green line). The extracted $A_{\rm g}$ (the value is indicated) is used to plot ${\cal D}_{\rm G}(\omega)\!=\!A_{\rm g}\,\omega^4$ in panel (a). See text for discussion.}
\label{fig:swap_fig}
\end{figure}

To further demonstrate the validity of Eq.~\eqref{eq:additive}, and the power and merit of the excess modes enumeration scheme based on it, we conclude the paper with an analysis of the PIPL model. This computer glass former is tailored for the Swap-Monte-Carlo algorithm~\cite{LB_swap_prx} such that it can be deeply supercooled, down to very low equilibrium temperatures. We supercooled PIPL liquid states to a very low temperature $T_{\rm p}\!=\!0.32$ (in simulation units, see details in~\cite{boring_paper}), which are subsequently quenched instantaneously to zero temperature, forming glasses (hence, $T_{\rm p}$ is termed the `parent' temperature of the generated glass samples). 2000 independent samples of $N\!=\!16000$ particles were generated, constituting a relatively large statistical ensemble. Despite this, these ultrastable glasses feature a very small number of QLEs below the first phonon peak. This is manifested in the noisy low-frequency tail of the VDoS ${\cal D}(\omega)$ observed in Fig.~\ref{fig:swap_fig}a, which does not allow to robustly extract the prefactor $A_{\rm g}$ in ${\cal D}_{\rm G}(\omega)\=A_{\rm g}\,\omega^4$.

In Fig.~\ref{fig:swap_fig}b, similarly to Fig.~\ref{fig:extract_Ag}c and Fig.~\ref{fig:ipl_vdos}, we plot ${\cal C}(\omega)\!-\!{\cal C}_{\rm D}(\omega)$ and fit it to $A_{\rm g}\omega^5/5$. The fit is very good, yielding $A_{\rm g}\!\approx\!2.7\!\times\!10^{-6}$, which is used to superimpose ${\cal D}_{\rm G}(\omega)\=A_{\rm g}\,\omega^4$ in Fig.~\ref{fig:swap_fig}a (continuous green line). We conclude that the developed procedure provides a much more robust approach to enumerating nonphononic excitations in ultrastable glasses, compared to alternative approaches.

\section{Summary and outlook}
\label{sec:summary}

In this work, we established two major, interrelated results. First, we demonstrated that the total VDoS of glasses, ${\cal D}(\omega)$, emerges from the sum of phononic, ${\cal D}_{\rm D}(\omega)$, and nonphononic, ${\cal D}_{\rm G}(\omega)$, contributions according to Eq.~\eqref{eq:additive}. This basic result indicates that while phonons and QLEs strongly hybridize in space, hybridization involves modes of narrowly distributed vibrational frequencies near a given $\omega$ such that the respective number of each species contributes additively to the total VDoS per $\omega$. This result also shows that all low-frequency excess modes are QLEs, i.e., there are no other types of nonphononic excitations.

Second, based on the validity of the additive structure of Eq.~\eqref{eq:additive}, we developed a general, simple and
straightforward scheme to enumerate nonphononic modes in computer glasses. It only requires the computation of the total VDoS ${\cal D}(\omega)$ and of the elastic constants, from which Debye's phononic VDoS ${\cal D}_{\rm D}(\omega)$ is obtained. Its great advantage is that it is insensitive to the hybridization of phonons and QLEs in space, which was the main stumbling block for extracting the nonphononic VDoS ${\cal D}_{\rm G}(\omega)$ for a long time. The validity of Eq.~\eqref{eq:additive}, established here, also opens the way for related analyses of experimental spectra of glasses, as demonstrated very recently in~\cite{experimental_evidence_2024_arXiv}.

Another advantage of the developed approach is that it allows to delineate the $\sim\!\omega^4$ scaling regime of the nonphononic VDoS (it is now well-established that QLEs exist beyond the power-law scaling regime, up to and in the vicinity of the boson peak frequency, see Fig.~\ref{fig:introduction_fig} and~\cite{boson_peak_2d_jcp_2023}). In the two models studied, we estimate the upper limit frequency of the scaling regime, denoted $\omega_{\rm c}$, to be $\omega_{\rm c}\!\approx\!1.5$ in BIPL computer glasses (cf.~Fig.~\ref{fig:ipl_vdos}a), and $\omega_{\rm c}\!\approx\!2.0$ for PIPL computer glasses (cf.~Fig.~\ref{fig:swap_fig}b); interestingly, the ratios $\omega_{\rm c}/A_{\rm g}^{-1/5}\!\approx\!0.35$ for the BIPL model, and $\omega_{\rm c}/A_{\rm g}^{-1/5}\!\approx\!0.15$ for the PIPL model, indicating that this ratio decreases with increasing stability of the analyzed computer glasses (recall that $A_{\rm g}$ has the dimension of $[1/\mbox{frequency}]^5$). Future research should reveal how the ratio $\omega_{\rm c}/A_{\rm g}^{-1/5}$ depends on different features of other computer-glass models.

Our approach allows investigations of the nonphononic VDoS in larger computer glasses, which is important for mitigating finite-size effects. In particular, in small computer-glass samples quenched quickly from high parent temperature equilibrium states, ${\cal D}_G(\omega)\!\sim\!\omega^\beta$ is observed in the low-frequency tail with $\beta\!<\!4$~\cite{lerner2019finite,2d_spectra_jcp_2022,Xu_modes_nat_comm_2024}. The scheme presented here can substantially reduce these finite-size effects and allow for a more accurate determination of the prefactor $A_{\rm g}$ of the nonphononic spectrum, independently of the glass formation history. Finally, we note that one could take into account also the disorder-induced broadening of phonon bands~\cite{phonon_widths}, resulting in lifting their strict degeneracy and giving rise to a finite spectral width $\Delta\omega$, but this is unnecessary as far as the number of modes is concerned, as we demonstrated.\\

\acknowledgements

A.M.~acknowledges support from the James S.~McDonnell Foundation Postdoctoral Fellowship Award in Complex Systems (\url{https://doi.org/10.37717/2021-3362}). E.B.~acknowledges support from the Ben May Center for Chemical Theory and Computation and the Harold Perlman Family.


\begin{thebibliography}{39}
\bibitem [{foo({\natexlab{b}})}]{footnote3}%
  \BibitemOpen
  \bibinfo {note} {Debye's frequency in 2D takes the form $\omega_{\rm
  D}\!\equiv\!\sqrt{\frac{8\pi (N/V)}{c_{\rm s}^{-2} + c_{\rm \ell}^2}}$, where
  $N$ denotes the system size, $V$ the volume, and $c_{{\rm s},\ell}$ are the
  shear and sound wave-speeds, respectively.}\BibitemShut {Stop}%
\bibitem [{foo({\natexlab{c}})}]{footnote2}%
  \BibitemOpen
  \bibinfo {note} {QLEs can assume the form of non-hybridized harmonic
  vibrational modes only at frequencies
  $\omega\!<\!\omega_\dagger(L)\!\sim\!L^{-\frac{2}{\dbar+2}}$, in $\dbar$
  spatial dimensions~\cite{phonon_widths}.}\BibitemShut {Stop}%
\end{thebibliography}

%

\end{document}